\documentclass[12pt]{iopart}

\usepackage{iopams}

\usepackage{iopams}
\usepackage{amssymb}
\usepackage{graphicx}
\usepackage{dcolumn}
\usepackage{bm}
\usepackage{subfigure}
\usepackage{float}
\usepackage{color}
\usepackage{cite}
\begin{document}

\title[]{Towards implementation of a magic optical-dipole trap for confining ground-state and Rydberg-state cesium cold atoms}

\author{Jiandong Bai$^{1}$, Shuo Liu$^{1}$, Jun He$^{1,2}$, and Junmin Wang$^{1,2,*}$}

\address{$^{1}$State Key Laboratory of Quantum Optics and Quantum Optics Devices, and Institute of Opto-Electronics, Shanxi University, Tai Yuan 030006, P. R. China}
\address{$^{2}$Collaborative Innovation Center of Extreme Optics of the Ministry of Education and Shanxi Province, Shanxi University, Tai Yuan 030006, P. R. China}
\ead{wwjjmm@sxu.edu.cn (Junmin WANG)}
\vspace{10pt}
\begin{indented}
\item[]January 2020
\end{indented}

\begin{abstract}
Long ground-Rydberg coherence lifetime is interesting for implementing high-fidelity quantum logic gates, many-body physics, and other quantum information protocols. However, the potential formed by a conventional far-off-resonance red-detuned optical-dipole trap (ODT) is usually repulsive for Rydberg atoms, which will result in fast atom loss and low repetition rate of the experimental sequence. These issues can be addressed by a magic ODT. We performed the calculation of ODT's magic detuning for confinement of cesium ground state and Rydberg state with the same potential well. We used a sum-over-states method to calculate the dynamic polarizabilities of $6S_{1/2}$ ground state and highly-excited ($nS_{1/2}$ and $nP_{3/2}$) Rydberg state of cesium atoms, and identify corresponding magic detuning for optical wavelengths in the range of $850 - 2000$ nm. We estimated the trapping lifetime of cesium Rydberg atoms confined in the magic ODT by including different dissipative mechanisms. Furthermore, we have experimentally realized an 1879.43-nm single-frequency laser system with a watt-level output power for setting up the magic ODT for $6S_{1/2}$ ground-state and $84P_{3/2}$ Rydberg-state cesium cold atoms.
\end{abstract}
\maketitle
%
% Uncomment for keywords
%\vspace{2pc}
%\noindent{\it Keywords}: XXXXXX, YYYYYYYY, ZZZZZZZZZ
%
% Uncomment for Submitted to journal title message
%\submitto{\NJP}
%
% Uncomment if a separate title page is required
%\maketitle
%
% For two-column output uncomment the next line and choose [10pt] rather than [12pt] in the \documentclass declaration
%\ioptwocol
%

\section{Introduction}

Rydberg atom refers to the atom in a highly-excited state with principal quantum number $n>10$. Compared with a low-excited state, Rydberg atoms have the characteristics of a long radiative lifetime and a large electric-dipole moment, and is extremely sensitive to DC or AC external electric fields \cite{gallagher2005}. Such exotic properties result in strong, controllable, long-range interactions among Rydberg atoms at spacings of micrometer. So Rydberg atoms are ideal systems for studying many-body physics \cite{zeiher2016, bernien2017}, and is promising in the fields of quantum information \cite{saffman2010}, quantum computing \cite{saffman2016}, nonlinear optics \cite{firstenberg2016}, electric field sensing \cite{sedlacek2013} and imaging \cite{gross2019}. In most of the experiments of cold atoms involving confinement of ground-state atoms in a far-off-resonance optical-dipole trap (ODT) \cite{grimm2000} and Rydberg excitation \cite{hankin2014,jau2016,urban2009,gaetan2009,arias2019}, cold atomic sample is prepared in an ODT to hold them in one position for a significant time. The potential formed by a conventional far-off-resonance red-detuned ODT is attractive for the ground-state atoms, but usually repulsive for highly-excited Rydberg-state atoms, leading that Rydberg atoms normally cannot be confined in the conventional ODT [Fig. 1(a)]. To preserve the long-lived coherence of the atoms, the magnetic trap \cite{choi2005} and electrostatic trap \cite{hyafil2004,hogan2008} for Rydberg atoms have been proposed and demonstrated. However, compared with the optical trap, these trap architectures are inflexible. The highly-excited Rydberg atom's size is generally larger than the constant of an optical lattice, so the nearly free valence electron will experience the ponderomotive potential. The optically trapping of rubidium (Rb) Rydberg atoms in a one-dimensional ponderomotive 1064-nm optical lattice was demonstrated by rapidly inverting the lattice potential \cite{anderson2011}. Recently, the three-dimensional trapping of $^{87}$Rb Rydberg states have been demonstrated in a hollow bottle beam trap generated by a spatial light modulator \cite{barredo2019}. To further extend the interaction time of Rydberg atoms, it is possibile to trap the long-lived circular Rydberg states in an ODT \cite{cortinas2019}. In the above mentioned experiments involving the trapping of Rydberg atoms, the following points should be addressed: (1) If switching off the ODT during Rydberg excitation and coherent manipulation, it will result in atomic dephasing due to the thermal diffusion of the atoms and the extremely low repetition rate of the experimental sequence; (2) If the ODT remains operation, it may cause a low excitation efficiency of Rydberg atoms as the transition frequency is spatially position-dependent on the excitation light; (3) When the Rydberg excitation light and the ODT light interact with the atoms, it may cause the photoionization of Rydberg atoms, which is secondary to atom loss by the thermal diffusion. If the ground-state atoms and the desired highly-excited Rydberg atoms experience the same potential, at least the above mentioned aspects (1) and (2) can be solved. It is very promising for high-fidelity two-qubit C-NOT gate operations \cite{urban2009,gaetan2009} and quantum simulation based on many-body physical system \cite{Labuhn2016}.

To eliminate the differential AC Stark shift of the desired transitions of atoms confined in an ODT, the so-called magic wavelength was proposed in Ref. \cite{katori1999} and experimentally demonstrated in strong-coupling cavity quantum electrodynamics system \cite{mckeever2003} for applications in quantum measurement and precision metrology \cite{ye2008}. Previous studies on the magic-wavelength ODT of neutral atoms mainly focus on the transition from ground state to a low-excited state \cite{mckeever2003,ye2008,wjm2014,notermans2014,safronova2016,liu2017}. In 2005, a blue-detuned magic-wavelength bottle beam trap was proposed for trapping the $5S_{1/2}$ ground state and $50D_{5/2}$ Rydberg state of Rb atoms \cite{saffman2005}, and recently Barredo \emph{et al.} experimentally demonstrated the bottle beam trap for Rydberg-state Rb atoms \cite{barredo2019}. However, the technique to generate the blue-detuned bottle-beam dark trap is generally complicated and not easy to implement \cite{barredo2019,xu2010}. In recent years, some progress has been made in the studies of magic ODT for Rydberg-state atoms, both theoretically \cite{saffman2005,goldschmidt2015} and experimentally \cite{li2013,wilson2019}. An one-dimensional pondermotive magic optical lattice for confining $5S_{1/2}$ ground-state and $90S_{1/2}$ Rydberg-state ${^{87}}$Rb atoms was demonstrated for the light-atom entanglement \cite{li2013}. The trapping of Rydberg-state ytterbium atoms also have been demonstrated in the same red-detuned optical tweezer that also confines the ground state \cite{wilson2019}. However, the research of the ODT's magic conditions for highly-excited Rydberg-state cesium (Cs) atoms is rare, especially for \emph{nP} Rydberg states.

In this paper, we calculate the dynamic polarizabilities of Cs $6S_{1/2}$ ground state, and Cs $nS_{1/2}$ and $nP_{3/2}$ Rydberg states and identify the corresponding magic conditions between 850 and 2000 nm. Generally there exists many magic conditions over a broad wavelength range. Compared with the magic ODT for low excited state, the magic ODT for Rydberg states is normally not so far-detuned, which ranges from several hundreds MHz to several GHz. Our goal is to identify the most useful magic detuning for Cs $nS_{1/2}$ and $nP_{3/2}$ Rydberg states \cite{hankin2014,jau2016,wjy2017,bjd2019,bjd2020}. The photon scattering rate in the ODT with a given trap potential depth is calculated and analyzed, and the trapping lifetime of Rydberg atoms confined in the magic ODT is discussed. For the magic ODT experiments, one need to produce a similar trap potential depth with sufficient laser power at the magic condition. Furthermore, the photon scattering rate should be low to preserve the coherence between ground state and Rydberg state.

\section{Calculation methods}
\subsection{Quantum defect of Rydberg-state atoms}

For alkali metal Rydberg atoms, the valence electron is far away from the nucleus, so the inner electrons and the nucleus can be regarded as an ``atomic core'' with a positive charge, which can then be treated as hydrogen-like atom. There are two kinds interaction between the outermost electron and the atomic core: the penetration and polarization. For the energy state with low orbital angular momentum quantum number $l$, the penetration effect plays a major role in reduction (so called quantum defect) of the valence electron energy. When the electron penetrates into the atomic core, they are affected by the nuclear charge originally shielded by the inner electrons, and the binding energy increases and the total energy decreases. For the high-$l$ states, the outermost electron almost no longer penetrate the atomic core due to the centrifugal potential $l(l+1)/{{2r}^{2}}$, and only the polarization contributes to reduction of the valence electron energy. For the low-$l$ states, considering the polarization and orbital penetration effect of Rydberg atoms, the binding energy of electron can be expressed as following \cite{gallagher2005}
%Eqn{1}
\begin{eqnarray}
E_{n,l,j} &=& E_{IP}-\frac{R_{Cs}}{{(n-\delta_{n,l,j})}^{2}}
\end{eqnarray}
where ${{E}_{IP}}$ = 31406.467669 ${{cm}^{-1}}$ is the ionization threshold energy, ${{R}_{Cs}}$ = 109736.86274 ${{cm}^{-1}}$ is the Rydberg constant for Cs atoms. The quantum defect factor ${\delta_{n,l,j}}$ is given by
%Eqn{2}
\begin{eqnarray}
\delta_{n,l,j} &=& \delta_{0}+\frac{\delta_{2}}{{(n-\delta_{0})}^{2}}+\frac{\delta_{4}}{{(n-\delta_{0})}^{4}}+\frac{\delta_{6}}{{(n-\delta_{0})}^{6}}+\cdots
\end{eqnarray}
where the expansion coefficients (${\delta_{0}}$, ${\delta_{2}}$, ${\delta_{4}}$, $\cdots$) are found by fitting the energy of the observed Rydberg spectra. These parameters are listed in Table 1, which are taken from Refs. \cite{weber1987,lorenzen1984}. For highly-excited Rydberg atoms, it is sufficient to consider only the first two terms in Eq. (2). We utilize Eqs. (1) and (2) to calculate the energies of Rydberg states, which are further used to calculate the atomic polarizability in the next section.

%Table{1}
\begin{table}[htbp]
\vspace{-0.0in}
\caption{The expansion coefficients of quantum defect for Cs atom in Eq. (2). These values are taken from Ref. \cite{weber1987,lorenzen1984}.}
\footnotesize
\begin{tabular*}{\textwidth}{@{}r*{15}{@{\extracolsep{0pt plus 12pt}}r}}
\br
	 & ${\delta_{0}}$ & ${\delta_{2}}$ & ${\delta_{4}}$ & ${\delta_{6}}$ & ${\delta_{8}}$ \\
\mr
	${nS}_{1/2}$ & 4.0493567 & 0.2377037 & 0.255401 & 0.00378 & 0.25486\\
	${nP}_{1/2}$ & 3.5915895 & 0.360926	& 0.41905	& 0.64388	& 1.45035\\
	${nP}_{3/2}$ & 3.5589599 & 0.392469	& -0.67431	& 22.3531	& -92.289\\
	${nD}_{3/2}$ & 2.4754562 & 0.00932	& -0.43498	& -0.76358	& -18.0061\\
	${nD}_{5/2}$ & 2.4663152 & 0.013577	& -0.37457	& -2.1867	& -1.5532\\
\br
	\end{tabular*}
\end{table}

\subsection{Atomic polarizability}
The polarizability of the atomic state $\alpha(\omega)$ is composed of a contribution from the valence electron $\alpha^{v}(\omega)$ and a core polarizability. The contribution of valence electron to the scalar and tensor polarizabilities can be calculated by summing over the intermediate states ($\vert n'l'j'\rangle$) that satisfy the electric-dipole transition selection rules. The scalar polarizability determines the frequency shift of energy level, and the tensor polarizability determines the splitting of energy level. According to the theory of the quadratic Stark effect, the dynamic scalar polarizability $\alpha_0^v (\omega)$ and tensor polarizability $\alpha_2^v(\omega)$ are expressed as following \cite{khadjavi1968}
%Eqn{3}
\begin{eqnarray}
{\alpha_0^v}(\omega) &= \frac{2}{3(2j+1)} \sum_{n'l'j'} \frac{{\vert\langle n'l'j'\parallel er\parallel nlj\rangle\vert}^2 (E_{n'l'j'}-E_{nlj})}{(E_{n'l'j'}-E_{nlj})^2-\omega^2}
\end{eqnarray}

%Eqn{4} 
\begin{eqnarray} 
{\alpha_2^v}(\omega) &= \sqrt{\frac{40j(2j-1)}{3(j+1)(2j+1)(2j+3)}} \sum_{n'l'j'} (-1)^{j+j'} 
\nonumber \\ & \times  {\left\{\begin{array}{ccc} j & j' & 1\\ 1 & 2 & j\end{array}\right\}} \frac{{\vert\langle n'l'j'\parallel er\parallel nlj\rangle\vert}^2 (E_{n'l'j'}-E_{nlj})}{(E_{n'l'j'}-E_{nlj})^2-\omega^2}
\end{eqnarray}
where the transition matrix element is expressed as following
%Eqn{5}
\begin{eqnarray}
{\vert\langle n'l'j'\parallel er\parallel nlj\rangle\vert}^2 &= (2j+1)(2j'+1) {\left\{\begin{array}{ccc} l & j & \frac{1}{2} \nonumber \\ j' & l' & 1\end{array}\right\}}^2 \\ &\times max(l,l') {\vert\langle n'l'\parallel er\parallel nl\rangle\vert}^2 
\end{eqnarray}
here $\langle n'l'\parallel er\parallel nl\rangle$ represents the radial electric-dipole matrix element, $max(l, l')$ is the larger number in $l$ and $l'$, and $\{ \cdots \}$ represents Wigner $6j$ coefficients. The core polarizability is almost independent of $\omega$, so it can be approximated by its static value \cite{mitroy2010}. The core polarizability of Cs atoms is 15.81  $a_0^3$ ($a_0$ is Bohr radius) \cite{mitroy2010}.

The total polarizability is given by
%Eqn{6}
\begin{eqnarray}
\alpha = \alpha_{core} + \alpha_0^v + \alpha_2^v \frac{3m_j^2 - j(j+1)}{j(2j-1)}
\end{eqnarray}
where $j$ is the total angular momentum quantum number, and $m_j$ is the corresponding magnetic quantum number. So the total polarizabilities for the $S_{1/2}$,  $P_{1/2}$, $P_{3/2}$, $D_{3/2}$, and $D_{5/2}$ states can be expressed as following
%Eqn{7}
\begin{eqnarray}
\alpha(S_{1/2}) &= \alpha(P_{1/2}) = \alpha_0 \nonumber \\
\alpha(P_{3/2}, \vert m_j\vert = 1/2) &= \alpha(D_{3/2}, \vert m_j\vert = 1/2) = \alpha_0 - \alpha_2^v \nonumber \\
\alpha(P_{3/2}, \vert m_j\vert = 3/2) &= \alpha(D_{3/2}, \vert m_j\vert = 3/2) = \alpha_0 + \alpha_2^v \nonumber\\
\alpha(D_{5/2}, \vert m_j\vert = 1/2) &= \alpha_0 - \frac{4}{5} \alpha_2^v \\
\alpha(D_{5/2}, \vert m_j\vert = 3/2) &= \alpha_0 - \frac{1}{5} \alpha_2^v \nonumber \\
\alpha(D_{5/2}, \vert m_j\vert = 5/2) &= \alpha_0 + \alpha_2^v \nonumber
\end{eqnarray}
here, $\alpha_0$ = $\alpha_{core}$ + $\alpha_0^v$ is the scalar part of the total polarizability. For the $j$ = 1/2 ($S_{1/2}$ and $P_{1/2}$) state, there is no tensor polarizability. However, for the $j>$ 1/2 ($P_{3/2}$, $D_{3/2}$, and $D_{5/2}$) state, the total polarizability is determined by both scalar and tensor polarizabilities, which depends upon $m_j$. So the magic condition need to be determined separately for the cases with $\vert m_j \vert$ = 1/2, 3/2, $\cdots$, $j$ for the $6S_{1/2} \leftrightarrow nl$ transitions, owing to the tensor contribution to the total polarizability of the $j>$ 1/2 state.

\section{Magic ODT for ground-state and Rydberg-state Cs atoms}
\subsection{Magic ODT for $6S_{1/2}$ ground-state and $43S_{1/2}$ Rydberg-state Cs atoms}

The cascade two-photon excitation scheme from the ground state $\vert g\rangle$ = $\vert 6S_{1/2}\rangle$ to Rydberg state $\vert r\rangle$ = $\vert 43S_{1/2}\rangle$ via an intermediate state $\vert e\rangle$ = $\vert 6P_{3/2}\rangle$ coupled by the 852-nm and 510-nm lasers is shown in Fig. 1(b). The dynamic polarizabilities are evaluated by the same way as the static values, but setting $\omega \neq$ 0. In this paper, all calculations are performed for linear polarization. We give the dynamic polarizabilities of Cs $6S_{1/2}$ and $43S_{1/2}$ states for optical wavelengths from 850 to 2000 nm in Fig. 2(a). The calculation of the polarizability of $6S_{1/2}$ state is similar to Ref. \cite{wjm2014}. The dynamic polarizability of the $43S_{1/2}$ state is given by

%Eqn{8}
\begin{eqnarray}
{\alpha_{43S}}(\omega) &= \frac{1}{3} \sum_{n} \frac{{\vert\langle nP_{1/2}\parallel er\parallel 43S_{1/2}\rangle\vert}^2(E_{nP_{1/2}}-E_{43S_{1/2}})}{(E_{nP_{1/2}}-E_{43S_{1/2}})^2-\omega^2}
&+ \alpha_{core} \nonumber \\
&+ \frac{1}{3} \sum_{n} \frac{{\vert\langle nP_{3/2}\parallel er\parallel 43S_{1/2}\rangle\vert}^2(E_{nP_{3/2}}-E_{43S_{1/2}})}{(E_{nP_{3/2}}-E_{43S_{1/2}})^2-\omega^2}
\end{eqnarray}

% FIG. 1
\begin{figure}[htbp]
\vspace{-0.00in}
\centerline{
\includegraphics[width = 120mm]{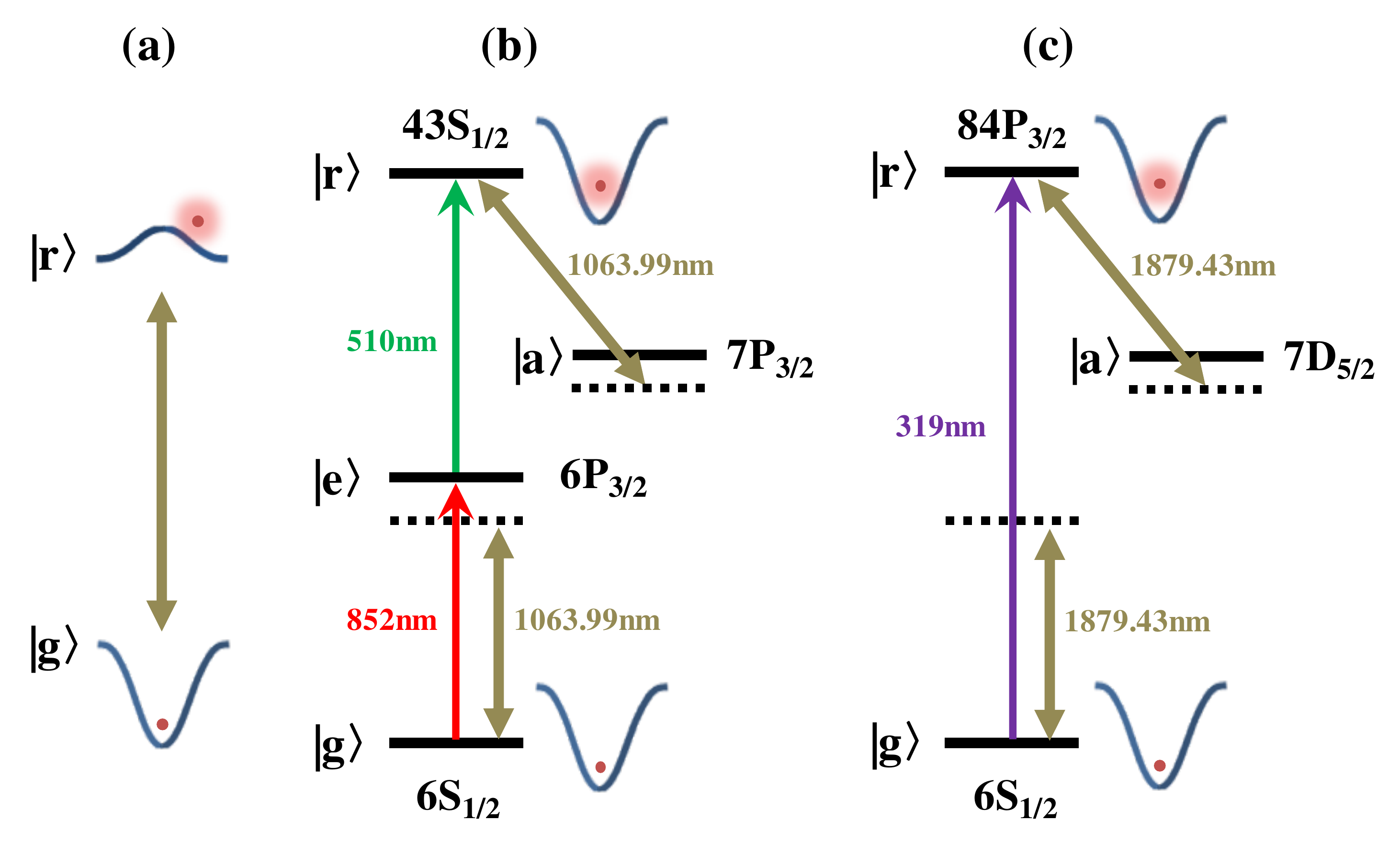}
} \vspace{-0.2in} %\setlength{\columnwidth}{3.2in}
\caption{Magic ODT for ground-state and Rydberg-state Cs atoms. (a) The far-off-resonance red-detuned ODT is attractive for ground states, but usually repulsive for highly-excited Rydberg states because almost all strong dipole transitions connected Rydberg state and the lower states have longer wavelength than that of ODT laser. (b) The cascade two-photon excitation scheme from $\vert g\rangle$ = $\vert 6S_{1/2}\rangle$ to $\vert r\rangle$ = $\vert 43S_{1/2}\rangle$ via $\vert e\rangle$ = $\vert 6P_{3/2}\rangle$ coupled by the 852-nm and 510-nm lasers. A 1063.99-nm laser is tuned to the blue side of the $\vert r\rangle \leftrightarrow \vert a\rangle $ =  $\vert 7P_{3/2}\rangle$ auxiliary transition to equalize the trapping potential depth of the  $\vert g\rangle$ and  $\vert r\rangle$ states. (c) The direct single-photon excitation scheme from $\vert g\rangle$ = $\vert 6S_{1/2}\rangle$ to $\vert r\rangle$ = $\vert 84P_{3/2}\rangle$ coupled by a 319-nm ultraviolet laser. An 1879.43-nm laser is also tuned to the blue side of the $\vert r\rangle \leftrightarrow \vert a\rangle $ =  $\vert 7D_{5/2}\rangle$ auxiliary transition to equalize the trapping potential depth of the $\vert g\rangle$ and  $\vert r\rangle$ states.}
\label{Fig 1}
\vspace{-0.1in}
\end{figure}

% FIG. 2
\begin{figure}[tbp]
\setlength{\belowcaptionskip}{-0.5cm}
\vspace{0in}
\centerline{
\includegraphics[width=160mm]{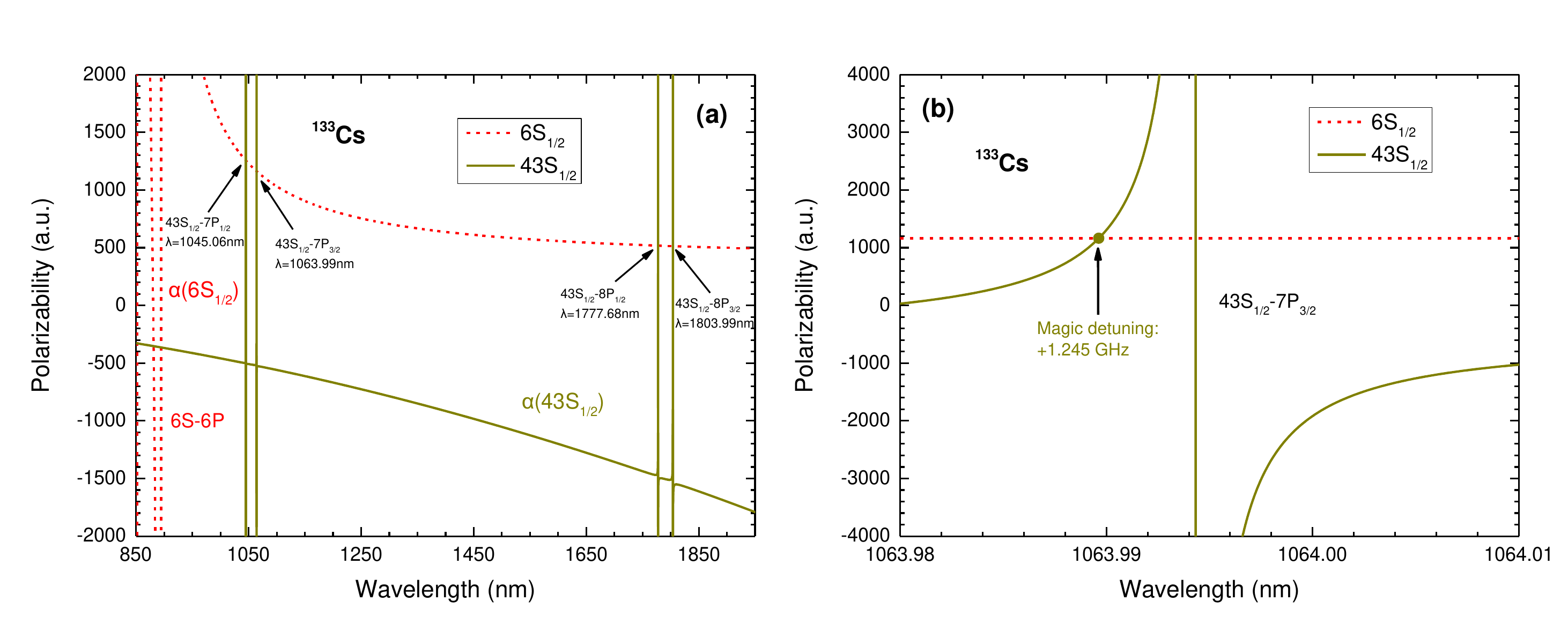}
} \vspace{-0.2in} %\setlength{\columnwidth}{3.2in}
%\centerline{
%}
\caption{(a) The dynamic polarizabilities of the $6S_{1/2}$ ground state (red dotted line) and the $43S_{1/2}$ Rydberg state (brown solid line) with the wavelength of the ODT between 850 and 2000 nm. (b) The dynamic polarizabilities near the $43S_{1/2} \leftrightarrow 7P_{3/2}$ auxiliary transition. If the polarizability is positive, the light shift is negative, leading to a potential well. When the detuning is +1.245 GHz relative to the transition of $43S_{1/2}(\vert m_j\vert = 1/2) \leftrightarrow 7P_{3/2}(\vert m_j\vert = 1/2)$, the ODT is attractive potential for the ground state and Rydberg state.}
\label{Fig 2}
\vspace{0.1in}
\end{figure}

%Table{2}
\newcommand{\tabincell}[2]{\begin{tabular}{@{}#1@{}}#2\end{tabular}} %在表格中自动换行
\begin{table}[htbp] %htbp为位置控制
\caption{Calculated magic detuning for $\vert r\rangle$ = $43S_{1/2} \leftrightarrow$ $\vert a\rangle$ auxiliary transition with the corresponding polarizability at the magic detuning. The other columns give the required optical intensity of the ODT laser with a trap potential depth of 1 mK and the wavelength of the nearest transition line.}
\footnotesize
\begin{tabular*}{\textwidth}{@{}l*{15}{@{\extracolsep{0pt plus12pt}}cr}}
\br
	Transition $\vert r\rangle \leftrightarrow \vert a\rangle$	& \tabincell{c}{Transition\\wavelength (nm)} & \tabincell{c}{$\alpha_{magic}$\\(a.u.)}	& \tabincell{c}{Magic\\detuning (MHz)} & \tabincell{c}{Optical\\intensity\\$(W/cm^2)$}\\\hline
	$43S_{1/2}(\vert m_j\vert = 1/2) \leftrightarrow 7P_{1/2}(\vert m_j\vert = 1/2)$ & 1045.06 & 1256.3 & +650 & $1.77\times 10^{5}$\\
	$43S_{1/2}(\vert m_j\vert = 1/2) \leftrightarrow 7P_{3/2}(\vert m_j\vert = 1/2)$ & 1063.99 & 1164.0 & +1245 & $1.91\times 10^{5}$\\
	$43S_{1/2}(\vert m_j\vert = 1/2) \leftrightarrow 8P_{1/2}(\vert m_j\vert = 1/2)$ & 1777.67 & 518.1 & +1488 & $4.29\times 10^{5}$\\
	$43S_{1/2}(\vert m_j\vert = 1/2) \leftrightarrow 8P_{3/2}(\vert m_j\vert = 1/2)$ & 1803.96 & 513.6 & +2586 & $4.33\times 10^{5}$\\
\br
	\end{tabular*}
\end{table}
\normalsize

%Table{3}
\begin{table}[h]
\caption{Contributions to the dynamic polarizability (in a.u.) of Cs $43S_{1/2}$ state at the magic detuning of +1.245 GHz relative to $43S_{1/2}(\vert m_j\vert = 1/2) \leftrightarrow 7P_{3/2}(\vert m_j\vert = 1/2)$ auxiliary transition. Reduced electric-dipole matrix elements (in a.u.) and the $43S_{1/2} \leftrightarrow nP$ energy differences (in GHz) are given in columns $D$ and $\Delta E$, respectively.}
\centering
\footnotesize
\begin{tabular}{lrrr}
\br
	Contribution & $D$ & $\Delta E $(GHz) & $\alpha $(a.u.)\\\hline
	${(6-38)P}_{1/2}$ &  &  & 41.2 \\
	${39P}_{1/2}$ & 42.515 & -455.581 & 22.7\\
	${40P}_{1/2}$ & 77.097 & -313.414 & 51.5\\
	${41P}_{1/2}$ & 188.127 & -182.495 & 178.4\\
	${42P}_{1/2}$ & 1259.406 & -61.669 & 2702.2\\
	${43P}_{1/2}$ & 1474.421 & 50.077 & -3007.4\\
	${44P}_{1/2}$ & 160.554 & 153.629 & -109.4\\
	${45P}_{1/2}$ & 62.705 & 249.769 & -27.1\\
	${46P}_{1/2}$ & 34.579 & 339.189 & -11.2\\
	${>46P}_{1/2}$ &   &   & -21.1 \\
 \\
	${6P}_{3/2}$ & 0.024	& -587648.094 & -0.003\\
	${7P}_{3/2}$ & 0.044	& -281761.332 & 1686.1 \\
	${(8-38)P}_{3/2}$ &    &             & 66.1\\
	${39P}_{3/2}$ & 54.482 & -450.756 & 37.0\\
	${40P}_{3/2}$ & 99.872 & -308.975 & 85.1\\
	${41P}_{3/2}$ & 249.787 & -178.402 & 307.5\\
	${42P}_{3/2}$ & 1865.609 & -57.887 & 5565.9\\
	${43P}_{3/2}$ & 2008.222 & 53.578 & -5969.4\\
	${44P}_{3/2}$ & 254.298 & 156.877 & -280.3\\
	${45P}_{3/2}$ & 103.236 & 252.788 & -74.4\\
	${46P}_{3/2}$ & 58.048 & 341.999 & -31.8\\
	${>46P}_{3/2}$ &       &         & -63.4\\
	core &       &         & 15.8\\
	Total &       &         & 1164.0\\
\br
	\end{tabular}
\end{table}

The positions of the $43S_{1/2} \leftrightarrow nP$ transitions are indicated by the arrow lines together with the corresponding $nP$ labels in Fig. 2. The magic detuning is determined as the crossing points of the polarizability curves of $6S_{1/2}$ and $43S_{1/2}$ states, which are listed in Table 2. There are four auxiliary transitions in the range of 850-2000 nm, corresponding to $43S_{1/2}(\vert m_j\vert = 1/2) \leftrightarrow 7P_{1/2,3/2}$, $8P_{1/2,3/2}$ $(\vert m_j\vert = 1/2)$ transitions, respectively. For $43S_{1/2}(\vert m_j\vert = 1/2) \leftrightarrow 7P_{3/2}(\vert m_j\vert = 1/2)$ auxiliary transition, it is very interesting due to commercial availability of the corresponding laser and high-power ytterbium-doped fiber amplifer (YDFA) near 1064 nm. So we give the dynamic polarizabilities of $6S_{1/2}$ and $43S_{1/2}$ states near 1064 nm, as shown in Fig. 2(b). The magic detuning relative to the auxiliary transition is +1.245 GHz. Therefore, to equalize the trapping potential depth of the $\vert g\rangle$ and  $\vert r\rangle$ states, a 1063.99-nm laser should be tuned to the blue side of $\vert r\rangle \leftrightarrow \vert a\rangle $ = $\vert 7P_{3/2}\rangle$ auxiliary transition, as shown in Fig. 1(b).

In order to verify the contributions of the neighboring intermediate states to the polarizability of $43S_{1/2}$ state, we use sum-over-states method \cite{mitroy2010} to perform calculations of polarizabilities. Calculated results are listed in Table 3, in which the various contributions to $\alpha_{43S}(\omega)$ are given at the magic detuning of +1.245 GHz relative to $43S_{1/2}(\vert m_j\vert = 1/2) \leftrightarrow 7P_{3/2}(\vert m_j\vert = 1/2)$ auxiliary transition. The contributions from 17 dominant transitions are listed separately with the corresponding reduced electric-dipole matrix elements $D$ and $43S_{1/2}\leftrightarrow nP$ energy difference $\Delta E$, respectively. The electric-dipole matrix elements are taken from Ref. \cite{sibalic2017}. These energy levels are calculated by using of Eqs. (1) and (2) in the section 2.1. We also compare the calculated value of the dynamic polarizability under the conditions of $\vert n-n'\vert \leq 2$ and $\vert n-n'\vert \leq 6$ including $43S_{1/2} \leftrightarrow 7P_{3/2}$ auxiliary transition, respectively. By comparison, we find that the errors between the two cases are in the range of 7\% and 2\%, respectively. Therefore, the calculation under the condition of $\vert n-n'\vert \leq 6$ including the nearest transition is sufficient for identifying magic detuning for confining $6S$ ground-state and $nl$ Rydberg-state Cs atoms.

\subsection{Magic ODT for $6S_{1/2}$ ground-state and $84P_{3/2}$ Rydberg-state Cs atoms}

The single-photon excitation scheme from ground state $\vert g\rangle$ = $\vert 6S_{1/2}\rangle$ to Rydberg state $\vert r\rangle$ = $\vert 84P_{3/2}\rangle$ coupled by a 319-nm ultraviolet (UV) laser is shown in Fig. 1(c). Compared to the cascade two-photon excitation scheme [Fig. 1(b)], the single-photon Rydberg excitation scheme has unique advantages: firstly, it can avoid the atomic decoherence due to population of atoms in the intermediate state in the mutli-photon excitation scheme. Although the photon scattering can be weaken by increasing the detuning of excitation light relative to the transition in every step, it cannot be eliminated completely. In particular, when the Rydberg-dressing atoms are used to generate tunable long-lived many-body interactions, the photon scattering from intermediate states in the multi-photon excitation scheme can not be negligible. Secondly, it can avoid the atomic dephasing and light shifts of ground and Rydberg states induced by the upper and lower excitation laser beams in the cascade multi-photon excitation scheme.

% FIG. 3
\begin{figure}[tp]
\setlength{\belowcaptionskip}{0.0cm}
\vspace{-0.0in}
\centerline{
\includegraphics[width=160mm]{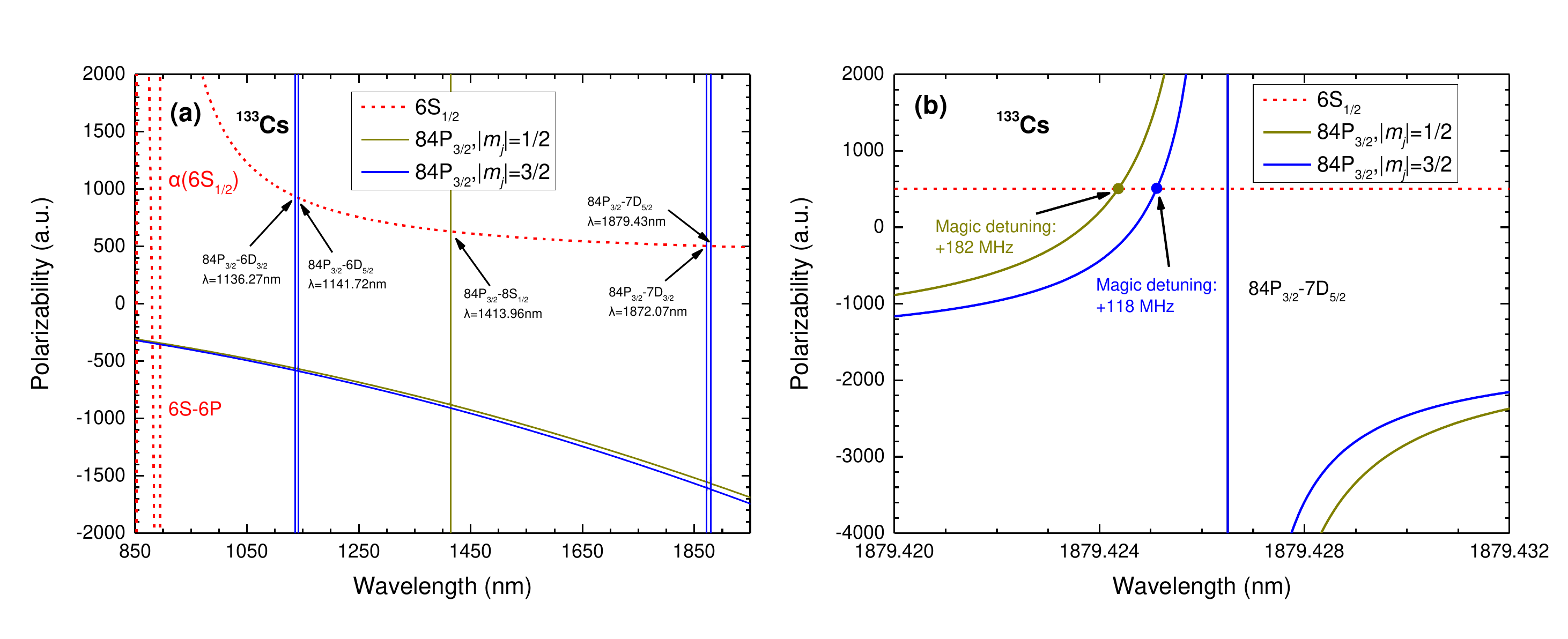}
} \vspace{-0.2in} %\setlength{\columnwidth}{3.2in}
%\centerline{
%}
\caption{(a) The dynamic polarizabilities of $6S_{1/2}$ ground state (dotted line) and highly-excited $84P_{3/2}$ Rydberg state (solid line) with wavelength of the ODT laser between 850 and 2000 nm. (b) The dynamic polarizabilities near the $84P_{3/2} \leftrightarrow 7D_{5/2}$ auxiliary transition. The results indicate that a focused single-spatial-mode Gaussian beam with the detuning of +182 MHz (or +118 MHz) relative to $84P_{3/2} \leftrightarrow 7D_{5/2}$ auxiliary transition can be served as the magic ODT to confine both $6S_{1/2}$- and $84P_{3/2}$-state Cs atoms with the same trap potential depth.}
\label{Fig 3}
\vspace{0.0in}
\end{figure}

%Table{4}
\begin{table*}[htbp] %htbp为位置控制
\caption{Calculated magic detuning for $\vert r\rangle$ = $84P_{3/2} \leftrightarrow$ $\vert a\rangle$ auxiliary transition with the corresponding polarizability. The other columns give the required intensity of ODT laser with a trap potential depth of 1 mK and wavelength of the nearest transition.}
\centering
\footnotesize
\begin{tabular*}{\textwidth}{@{}l*{15}{@{\extracolsep{0pt plus12pt}}cccc}}
\br
	Transition $\vert r\rangle \leftrightarrow \vert a\rangle$	& \tabincell{c}{Transition\\wavelength (nm)} & \tabincell{c}{$\alpha_{magic}$\\(a.u.)}	& \tabincell{c}{Magic\\detuning\\(MHz)} & \tabincell{c}{Optical\\intensity\\ $(W/cm^2)$}\\\hline
	$84P_{3/2}(\vert m_j\vert = 1/2) \leftrightarrow 6D_{3/2}(\vert m_j\vert = 1/2)$ & 1136.27 & 935.0 & $<$1 & $2.38\times 10^{5}$ \\
	$84P_{3/2}(\vert m_j\vert = 3/2) \leftrightarrow 6D_{3/2}(\vert m_j\vert = 3/2)$ & 1136.27 & 935.0 & +9 & $2.38\times 10^{5}$  \\
	$84P_{3/2}(\vert m_j\vert = 1/2) \leftrightarrow 6D_{5/2}(\vert m_j\vert = 1/2)$ & 1141.72 & 922.8 & +63 & $2.41\times 10^{5}$ \\
	$84P_{3/2}(\vert m_j\vert = 3/2) \leftrightarrow 6D_{5/2}(\vert m_j\vert = 3/2)$ & 1141.72 & 922.8 & +41 & $2.41\times 10^{5}$ \\
	$84P_{3/2}(\vert m_j\vert = 1/2) \leftrightarrow 8S_{1/2}(\vert m_j\vert = 1/2)$ & 1413.96 & 629.4 & +70 & $3.53\times 10^{5}$ \\
	$84P_{3/2}(\vert m_j\vert = 1/2) \leftrightarrow 7D_{3/2}(\vert m_j\vert = 1/2)$ & 1872.07 & 503.2 & +3 & $4.42\times 10^{5}$ \\
	$84P_{3/2}(\vert m_j\vert = 3/2) \leftrightarrow 7D_{3/2}(\vert m_j\vert = 3/2)$ & 1872.07 & 503.2 & +28 & $4.42\times 10^{5}$ \\
	$84P_{3/2}(\vert m_j\vert = 1/2) \leftrightarrow 7D_{5/2}(\vert m_j\vert = 1/2)$ & 1879.43 & 502.1 & +182 & $4.43\times 10^{5}$ \\
	$84P_{3/2}(\vert m_j\vert = 3/2) \leftrightarrow 7D_{5/2}(\vert m_j\vert = 3/2)$ & 1879.43 & 502.1 & +118 & $4.43\times 10^{5}$ \\
\br
	\end{tabular*}
\end{table*}

Although realization of UV laser for the single-photon Rydberg excitation is technically challenging, it is still of great significance for solving the physical issues. With the well-developed fiber laser, fiber amplifier and efficient frequency conversion PPXX (periodically poled nonlinear crystal) material and technology, we have implemented a narrow-linewidth continuous-tunable 319-nm UV laser with more than 2 W output \cite{wjy2016} for Cs $6S_{1/2} \leftrightarrow nP_{3/2} (n = 70-100)$ single-photon Rydberg excitation \cite{wjy2017,bjd2019,bjd2020}. To further realize the spatial localization of cold atomic ensemble against the atomic diffusion and decoherence caused by residual atomic thermal motion, a magic ODT will be a powerful tool for manipulating the cold atoms.

To implement a magic ODT for confining Cs $6S_{1/2}$ ground state and $nP_{3/2}$ Rydberg state, we use Eqs. (3)-(6) to calculate the dynamic polarizabilities of $6S_{1/2}$ and $84P_{3/2}$ states for $\lambda$ between 850-2000 nm, and results are plotted in Fig. 3(a). The auxiliary transitions are marked by arrows in the figure, corresponding to $84P_{3/2} \leftrightarrow 6D_{3/2,5/2}$, $8S_{1/2}$, and $7D_{3/2,5/2}$ transitions, respectively. The magic ODT near these auxiliary transitions is attractive potential for both ground and Rydberg states. Taking $84P_{3/2} \leftrightarrow 7D_{5/2}$ auxiliary transition as an example, a plot for the near resonant region is shown in Fig. 3(b). The polarizability curves of $84P_{3/2}$ state intersect with that of $6S_{1/2}$ ground state at two points, corresponding to $84P_{3/2} (\vert m_j\vert = 1/2) \leftrightarrow 7D_{5/2} (\vert m_j\vert = 1/2)$ and $84P_{3/2} (\vert m_j\vert = 3/2) \leftrightarrow 7D_{5/2} (\vert m_j\vert = 3/2)$ transitions, respectively. The magic detuning relative to the auxiliary transition are +182 and +118 MHz, respectively. The polarizability is 502.1 $a_0^3$ at the magic detuning to $84P_{3/2} \leftrightarrow 7D_{5/2}$ auxiliary transition. For an ODT with a trap potential depth of $\sim$ 1 mK, the required optical intensity is $4.43\times 10^5$ $W/cm^2$, which is listed in Table 4. Besides, Table 4 also presents the magic detuning for $84P_{3/2} \leftrightarrow 6D_{3/2,5/2}$, $8S_{1/2}$, and $7D_{3/2,5/2}$ auxiliary transitions. Compared to other auxiliary transitions, the magic detuning of +182 MHz to $84P_{3/2} (\vert m_j\vert = 1/2) \leftrightarrow 7D_{5/2} (\vert m_j\vert = 1/2)$ auxiliary transition is largest, so the photon scattering rate induced by the magic ODT laser beam with the same trap potential depth is lower.

\subsection{Trap potential of magic ODT}

A detailed description about ODTs can be found in Ref. \cite{grimm2000}. The trap potential depth $U$ of the ODT is expressed as following
%Eqn{9}
\begin{eqnarray}
U &= -\frac{I(r)}{2\epsilon_0 c} \alpha(\omega)
\end{eqnarray}
where $\epsilon_0$ is the permittivity of vacuum, $c$ is the speed of light, $\alpha(\omega)$ is the atomic polarizability, and $I(r)$ is the intensity profile of ODT laser beam. For the magic ODT for Cs $6S_{1/2}$ and $43S_{1/2}$ states, the polarizability is 1164.0 $a_0^3$ at the magic detuning for $43S_{1/2} (\vert m_j\vert = 1/2) \leftrightarrow 7P_{3/2} (\vert m_j\vert = 1/2)$ auxiliary transition at 1063.99 nm. When the trap potential depth is $\sim$ 1 mK, the ODT laser intensity is required to be approximately $1.91\times 10^5$ $W/cm^2$, which is listed in Table 2. Figure 2(a) shows that the polarizability of $6S_{1/2}$ ground state decreases with increase of wavelength of the ODT in the range of 900-2000 nm. When the polarizability decreases more power is needed to achieve a given trap depth. Therefore, the feasibility of the experiment becomes more difficult. For implementing the magic ODT, the 1063.99-nm single-frequency laser can be provided by a master oscillator power-amplifier (MOPA) laser system which is composed of a tunable distributed feedback diode laser at 1063.99 nm and a well-developed YDFA with a watt-level or even 10-watt-level output power. Moreover, it is also necessary to stabilize the laser frequency and control its detuning relative to the auxiliary transition with an ultra-low expansion (ULE) optical cavity. Considering the commercial high-power 1064-nm laser source and large polarizability of $43S_{1/2}$ state near 1064 nm, the magic ODT near 1064 nm is promising to confine both $6S_{1/2}$ ground-state and $43S_{1/2}$ Rydberg-state Cs atoms.

%Table{5}
\begin{table*}[htbp] %htbp为位置控制
\caption{Calculated magic detuning for $nS_{1/2}(\vert m_j\vert = 1/2) \leftrightarrow 7P_{3/2}(\vert m_j\vert = 1/2)$ auxiliary transition with the corresponding polarizability. The other columns give the required intensity of the ODT laser with a trap potential depth of 1 mK, and the orbital radius and the radiative lifetime of $nS_{1/2}$ Rydberg state at T = 300 K.}
\centering
\footnotesize
\begin{tabular*}{\textwidth}{@{}l*{15}{@{\extracolsep{0pt plus12pt}}cccccr}}
\br
	$n$ & \tabincell{c}{Orbital radius \\(nm)} & \tabincell{c}{Lifetime of \\$nS_{1/2}$ state($\mu$s)} & \tabincell{c}{Transition\\wavelength (nm)} & \tabincell{c}{$\alpha_{magic}$\\(a.u.)}	& \tabincell{c}{Magic detuning\\(MHz)} & \tabincell{c}{Optical\\intensity\\$(W/cm^2)$}\\\hline
	40 & 103 & 30 & 1065.42 & 1157.8 & +1592 & $1.92\times 10^{5}$\\
	43 & 120 & 37 & 1063.99 & 1164.0 & +1245 & $1.91\times 10^{5}$\\
	50 & 168 & 55 & 1061.69 & 1174.2 & +748 & $1.89\times 10^{5}$\\
	60 & 248 & 89 & 1059.79 & 1182.8 & +411 & $1.88\times 10^{5}$\\
	70 & 345 & 132 & 1058.69 & 1187.9 & +249 & $1.87\times 10^{5}$\\
	80 & 458 & 184 & 1057.99 & 1191.2 & +163 & $1.87\times 10^{5}$\\
	90 & 586 & 246 & 1057.53 & 1193.3 & +113 & $1.87\times 10^{5}$\\
	100 & 731 & 316 & 1057.20 & 1194.9 & +80 & $1.86\times 10^{5}$\\
\br
	\end{tabular*}
\end{table*}

%Table{6}
\begin{table*}[htbp] %htbp为位置控制
\caption{Calculated magic detuning for $nP_{3/2}(\vert m_j\vert = 1/2) \leftrightarrow 7D_{5/2}(\vert m_j\vert = 1/2)$ auxiliary transition  with the corresponding polarizability at the magic detuning. The other columns give the required intensity of ODT laser beam with a trap potential depth of 1 mK, and the orbital radius and the radiative lifetime of $nP_{3/2}$ Rydberg state at T = 300 K.}
\centering
\footnotesize
\begin{tabular*}{\textwidth}{@{}l*{15}{@{\extracolsep{0pt plus12pt}}cccccr}}
\br
	$n$ & \tabincell{c}{Orbital\\ radius (nm)} & \tabincell{c}{Lifetime of \\$nP_{3/2}$ state ($\mu$s)} & \tabincell{c}{Auxiliary transition\\wavelength (nm)} & \tabincell{c}{$\alpha_{magic}$\\(a.u.)} & \tabincell{c}{Magic detuning\\(MHz)} & \tabincell{c}{Optical\\intensity\\ $(W/cm^2)$}\\\hline
	40 & 105 & 47 & 1902.92 & 498.9 & +2013 & $4.45\times 10^{5}$ \\
	50 & 171 & 81 & 1891.49 & 500.5 & +954 & $4.44\times 10^{5}$ \\
	60 & 253 & 125 & 1885.62 & 501.3 & +526 & $4.43\times 10^{5}$ \\
	70 & 350 & 178 & 1882.22 & 501.8 & +322 & $4.43\times 10^{5}$ \\
	80 & 464 & 242 & 1880.07 & 502.0 & +211 & $4.43\times 10^{5}$ \\
	84 & 514 & 270 & 1879.43 & 502.1 & +182 & $4.43\times 10^{5}$ \\
	90 & 593 & 314 & 1878.62 & 502.3 & +147 & $4.43\times 10^{5}$ \\
	100 & 738 & 399 & 1877.61 & 502.4 & +106 & $4.42\times 10^{5}$\\
\br
	\end{tabular*}
\end{table*}

In addition, in the experiments involving trapping of ground-state atoms by an ODT and Rydberg excitation of the atoms, it is very interesting for most research groups to also confine Rydberg atoms in ODT. Therefore, we calculate the magic condition to trap both $6S_{1/2}$ ground state and $nS_{1/2}$ (n=40-100) Rydberg-state Cs atoms for $nS_{1/2} (\vert m_j\vert = 1/2) \leftrightarrow 7P_{3/2} (\vert m_j\vert = 1/2)$ auxiliary transition, including transition wavelength, polarizability, magic detuning, and the required intensity of the ODT laser, which are listed in Table 5. Moreover, these magic ODTs are relatively easy to be implemented due to the commercial YDFA with watt-level output power in wavelength range of 1030-1080 nm.

Furthermore, due to the availability of MOPA laser system with a watt-level output power at 1.9 $\mu$m, we also calculated the magic condition to confine both Cs $6S_{1/2}$ ground state and $nP_{3/2}$ (n=40-100) Rydberg state, including the polarizability, the required intensity of the magic ODT laser beam, and the magic detuning, which are listed in Table 6. It will provide useful references for those considering the magic ODT of Cs $nP_{3/2}$ Rydberg atoms.

\section{Photon scattering rate and trapping lifetime of Cs Rydberg state}

Compared with conventional ODT, the magic ODT for ground-state and Rydberg-state atoms only can be found around an auxiliary transition of $\vert r\rangle \leftrightarrow \vert a\rangle $, as shown in Fig. 1. Thus, photon scattering by the ODT laser beam will cause the atoms to be heated and quickly decoherent. In order to estimate the trapping lifetime of atoms in the magic ODT due to photon scattering, we calculate the photon scattering rate $R_{sc}$, which is expressed as following \cite{corney1977}
%Eqn{10}
\begin{eqnarray}
R_{sc} = \frac{\Gamma}{2}\frac{I/I_{sat}}{1+4(\Delta /\Gamma)^2+(I/I_{sat})}
\end{eqnarray}
here $I$ and $I_{sat}$ are the intensity of the ODT laser and the saturated light intensity, respectively. $\Delta$ is the detuning of the ODT laser relative to the auxiliary transition. The spontaneous decay rate $\Gamma$ (Einstein A coefficient) from $\vert n'j'\rangle$ to $\vert nj\rangle$, is related to oscillator strength $f$ by \cite{mitroy2010}
%Eqn{11}
\begin{eqnarray}
A_{n'j'nj} = \frac{e^2 \omega_0^2}{2\pi\epsilon_0 m_e c^3} \frac{2j+1}{2j'+1} f
\end{eqnarray}
for $j\rightarrow j'$ transition. Here, $e$ is the elementary charge, $m_e$ is the electron mass, $\epsilon_0$ is the permittivity of vacuum, and $c$ is the speed of light. The oscillator strengths can be expressed as following
%Eqn{12}
\begin{eqnarray}
f_{njn'j'} &= \frac{2m_e \omega_0}{3\hbar e^2} \frac{1}{2j+1}\vert\langle n'j'\parallel \bm{d} \parallel nj\rangle\vert^2
\end{eqnarray}
where $\hbar$ is the reduced Planck constant, $\omega_0$ = $(E_{n'l'j'}-E_{nlj})/{\hbar}$ is the transition angular frequency, and $\langle n'j'\parallel \bm{d}\parallel nj\rangle$ is the reduced matrix element.

Here we analyze the dissipative mechanisms which affect the decoherence time between ground and Rydberg states. The effective population decay lifetime of Rydberg state, $\tau_{eff}$ is given by \cite{gallagher2005}
%Eqn{13}
\begin{eqnarray}
\frac{1}{\tau_{eff}} = \frac{1}{\tau_{sc}} + \frac{1}{\tau_n^{(0)}} + \frac{1}{\tau_n^{(bb)}}
\end{eqnarray}
where $\tau_{sc}$, $\tau_n^{(0)}$, and $\tau_n^{(bb)}$ are the contributions from the ODT-induced photon scattering calculated by Eq. (10), vacuum induced radiative decay, and blackbody radiation at finite temperature, respectively. $\tau_{sc}$, $\tau_n^{(0)}$, and $\tau_n^{(bb)}$ are given by \cite{gallagher2005}
%Eqn{14}
\begin{eqnarray}
\tau_{sc} &= \frac{1}{R_{sc}}\nonumber \\
\tau_n^{(0)} &= \tau^{(0)} (n^*)^a \\
\tau_n^{(bb)} &= \frac{3\hbar (n^*)^2}{4\alpha_{fs}^3 k_B T} \nonumber
\end{eqnarray}
where $\tau^{(0)}$ = 1.43 and 4.42 ns, $a$ = 2.96 and 2.94 for $S$ and $P$ states of Cs atoms, $n^*$ = $n$ - $\delta_{n,l,j}$ is the effective principal quantum number, $\alpha_{fs}$ is the fine-structure constant, and $T$ = 300 K.

% FIG. 4
\begin{figure}[tp]
\setlength{\belowcaptionskip}{0.0cm}
\vspace{-0.0in}
\centerline{
\includegraphics[width=110mm]{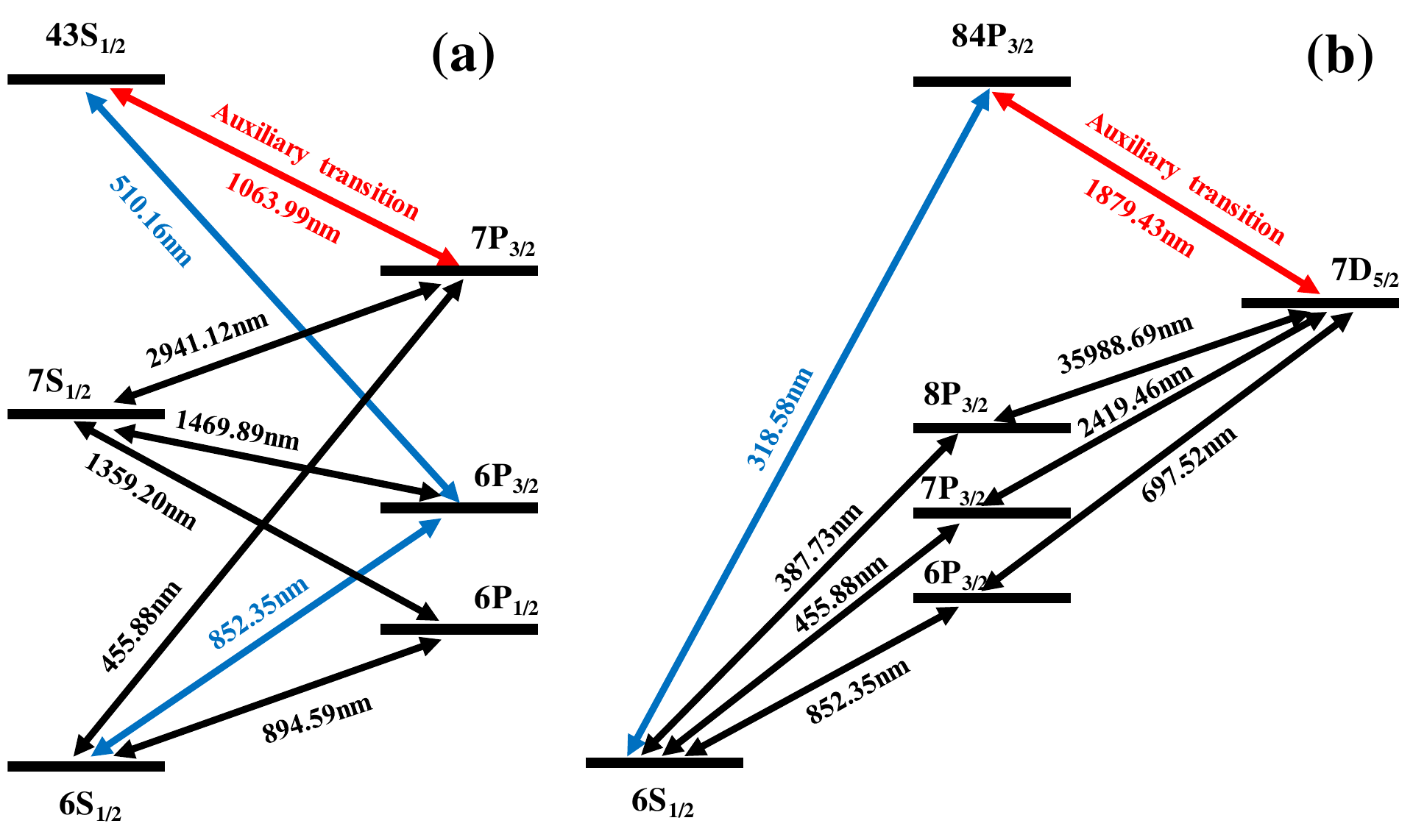}
} \vspace{-0.1in} %\setlength{\columnwidth}{3.2in}
%\centerline{
%}
\caption{Relevant energy levels and transitions for Cs $43S_{1/2}$ and $84P_{3/2}$ Rydberg states. (a) The $43S_{1/2}$ Rydberg state is prepared by cascade two-photon excitation (852.35 nm + 510.16 nm) scheme. A focused 1063.99-nm single-spatial-mode Gaussian laser beam near $43S_{1/2} \leftrightarrow 7P_{3/2}$ auxiliary transition is used to confine both $6S_{1/2}$ ground-state and $43S_{1/2}$ Rydberg-state atoms. (b)  The $84P_{3/2}$ Rydberg state can be prepared by direct 319-nm UV single-photon excitation scheme. A focused 1879.43-nm single-spatial-mode Gaussian laser beam near $84P_{3/2} \leftrightarrow 7D_{5/2}$ auxiliary transition is used to confine both $6S_{1/2}$ ground-state and $84P_{3/2}$ Rydberg-state atoms. }
\label{Fig 4}
\vspace{0.0in}
\end{figure}

For the magic ODT for Cs $6S_{1/2}$ ground state and $43S_{1/2}$ Rydberg state, the 1063.99-nm ODT laser's detuning relative to $43S_{1/2} \leftrightarrow 7P_{3/2}$ auxiliary transition is not so far, there is the mixing of $7P_{3/2}$ population that cause $7P_{3/2} \leftrightarrow 6S_{1/2}$ fluorescence photons to be produced, as shown in Fig. 4(a). So we estimate the photon scattering rate of both the auxiliary transition and the spontaneous emission from the auxiliary state $7P_{3/2}$, which are listed in Table 7. It shows that the photon scattering rate from the auxiliary transition is much more important. For the magic ODT for Cs $6S_{1/2}$ ground state and $84P_{3/2}$ Rydberg state, the 1879.43-nm ODT laser may yield weak population on $7D_{5/2}$ auxiliary state. The spontaneous emission from intermediate states to the $6S_{1/2}$ ground state involves these transitions, which are shown in Fig. 4(b). The photon scattering rates induced by the ODT laser are listed in Table 8.

%Table{7}
\begin{table*}[htbp] %htbp为位置控制
\caption{Photon scattering rates induced by the magic ODT laser at 1063.99 nm with an optical intensity of $1.91\times 10^{5}$ $W/cm^2$ (trap potential depth $\sim$ 1 mK) for $43S_{1/2} (\vert m_j\vert = 1/2) \leftrightarrow 7P_{3/2} (\vert m_j\vert = 1/2)$ auxiliary transition.}
\centering
\footnotesize
\begin{tabular*}{\textwidth}{@{}l*{15}{@{\extracolsep{0pt plus12pt}}rrrrcr}}
\br
	\tabincell{c}{Transition\\ $\vert r\rangle \leftrightarrow \vert a\rangle$} & \tabincell{c}{Transition\\ wavelength (nm)} & \tabincell{c}{Reduced matrix\\ element (e$a_0$)} & \tabincell{c}{Einstein A\\ coefficient $(s^{-1})$} & \tabincell{c}{Frequency\\ detuning\\ (GHz)} & \tabincell{c}{Photon\\ scattering\\ rate $(s^{-1})$}\\\hline
	$43S_{1/2} \leftrightarrow 7P_{3/2}$ & 1063.99 & 0.044 & $1.61\times 10^{3}$ & 1.245 & 19.0\\
	$43S_{1/2} \leftrightarrow 7P_{1/2}$ & 1045.06 & 0.032 & $9.24\times 10^{2}$ & -5102.484 & $3.6\times 10^{-7}$\\
	$7P_{3/2} \leftrightarrow 6S_{1/2}$ & 455.88 & 0.576 & $1.77\times 10^{6}$ & -375849.866 & $4.1\times 10^{-5}$\\
     $7P_{3/2} \leftrightarrow 7S_{1/2}$ & 2941.12 & 14.323 & $4.08\times 10^{6}$ & 179831.286 & 0.3\\
     $7S_{1/2} \leftrightarrow 6P_{3/2}$ & 1469.89 & 6.479 & $1.34\times 10^{7}$ & 77807.105 & 0.9\\
     $7S_{1/2} \leftrightarrow 6P_{1/2}$ & 1359.20 & 4.243 & $7.26\times 10^{6}$ & 61197.482 & 0.3\\
     $6P_{3/2} \leftrightarrow 6S_{1/2}$ & 852.35 & 6.324 & $3.27\times 10^{7}$ & -69963.104 & 2.6\\
     $6P_{1/2} \leftrightarrow 6S_{1/2}$ & 894.59 & 4.489 & $2.85\times 10^{7}$ & -53353.481 & 2.0\\
\br
	\end{tabular*}
\end{table*}

%Table{8}
\begin{table*}[htbp] %htbp为位置控制
\caption{Photon scattering rates induced by the magic ODT laser at 1879.43 nm with an optical intensity of $4.43\times 10^{5}$ $W/cm^2$ (trap potential depth $\sim$ 1 mK) for $84P_{3/2} (\vert m_j\vert = 1/2) \leftrightarrow 7D_{5/2} (\vert m_j\vert = 1/2)$ auxiliary transition.}
\centering
\footnotesize
\begin{tabular*}{\textwidth}{@{}l*{15}{@{\extracolsep{0pt plus12pt}}rrrrcr}}
\br
	\tabincell{c}{Transition\\ $\vert r\rangle \leftrightarrow \vert a\rangle$} & \tabincell{c}{Transition\\ wavelength (nm)} & \tabincell{c}{Reduced matrix\\ element (e$a_0$)} & \tabincell{c}{Einstein A\\ coefficient $(s^{-1})$} & \tabincell{c}{Frequency\\ detuning\\ (GHz)} & \tabincell{c}{Photon\\ scattering\\ rate $(s^{-1})$}\\\hline
	$84P_{3/2} \leftrightarrow 7D_{5/2} $ & 1879.43 & 0.024 & 43.51 & 0.182 & 7.0\\
	$84P_{3/2} \leftrightarrow 7D_{3/2} $ & 1872.07 & 0.008 & 4.63 & -626.565 & $1.6\times 10^{-9}$\\
	$84P_{3/2} \leftrightarrow 8S_{1/2} $ & 1413.96 & 0.004 & 3.58 & -52511.073 & $5.9\times 10^{-13}$\\
	$7D_{5/2} \leftrightarrow 6P_{3/2} $ & 697.52 & 2.890 & $8.31\times 10^{6}$ & -270282.471 & $1.3\times 10^{-2}$\\
	$7D_{5/2} \leftrightarrow 7P_{3/2} $ & 2419.46  & 9.600 & $2.20\times 10^{6}$ & 35604.290 & 2.2\\
	$7D_{5/2} \leftrightarrow 8P_{3/2} $ & 35988.69 & 43.200 & $1.35\times 10^{4}$ & 151182.728 & $1.5\times 10^{-2}$\\
	$6P_{3/2} \leftrightarrow 6S_{1/2} $ & 852.35	& 6.324 & $3.27\times 10^{7}$ & -192212.765 & 0.8\\
	$7P_{3/2} \leftrightarrow 6S_{1/2} $ & 455.88	& 0.576 & $1.77\times 10^{6}$ & -498099.527 & $5.4\times 10^{-5}$\\
	$8P_{3/2} \leftrightarrow 6S_{1/2} $ & 387.73	& 0.218 & $4.13\times 10^{5}$ & -613677.965 & $1.2\times 10^{-6}$\\
\br
	\end{tabular*}
\end{table*}

The photon scattering rates are not only determined by the frequency detuning, but also directly related to Einstein A coefficient of the auxiliary transitions. For Rydberg states, Einstein A coefficient is much smaller than that of the low-excited states. Therefore, even though the magic detuning is only hundreds of MHz relative to the auxiliary transition, the photon scattering rates are still not so high, as presented in Table 7 and Table 8. The scattering lifetime due to photon scattering is in the order of milliseconds, which is longer than that induced by both vacuum induced spontaneous decay and blackbody radiation when a specific auxiliary transition is chosen. Considering these dissipative mechanisms, lifetime of Cs $84P_{3/2}$ Rydberg atom in the magic trap is still close to its radiative lifetime at 300 K, which is extremely important for our subsequent experiment of direct single-photon Rydberg excitation. For Cs $43S_{1/2}$ Rydberg state, its trapping lifetime is approximately several tens microseconds by selecting a specific auxiliary transition.

\section{Magic ODT for Cs $6S_{1/2}$ ground state and $84P_{3/2}$ Rydberg state: Experimental consideration}

For Cs $84P_{3/2}$ Rydberg state, the experiment has been demonstrated with a direct 319-nm UV single-photon excitation \cite{hankin2014,wjy2017,bjd2020}. To spatially localize cold atomic ensemble and preserve the coherence between $6S_{1/2}$ ground state and $84P_{3/2}$ Rydberg state, we are implementing the magic ODT at 1879.43 nm to confine both the two states with the same trapping potential. And we have experimentally realized an 1879.43-nm laser system, as shown in Fig. 5(a). The 1879.43-nm single-frequency laser is produced by a MOPA laser system which is composed of a tunable single-frequency diode laser at 1879.43 nm and a 3-W custom-designed thulium-doped fiber amplifier (TmDFA) operating around 1879 nm. The efficiency of transferring Cs atoms from the magneto-optical trap to the ODT is severely restricted by polarization and intensity fluctuation of the ODT laser. A polarization controller is placed behind TmDFA to improve the polarization character. A Mach-Zehnder interferometer \cite{gao2019} is implemented with a feedback sero to actively suppress the laser intensity noise but with a fixed s polarization. At the present stage, the residual intensity fluctuation of 1879.43-nm laser beam is supressed from $\pm$ 6$\%$ to $\pm$ 0.6$\%$.

% FIG. 5
\begin{figure}[htpb]
\setlength{\belowcaptionskip}{0.0cm}
\vspace{-0.0in}
\centerline{
\includegraphics[width=140mm]{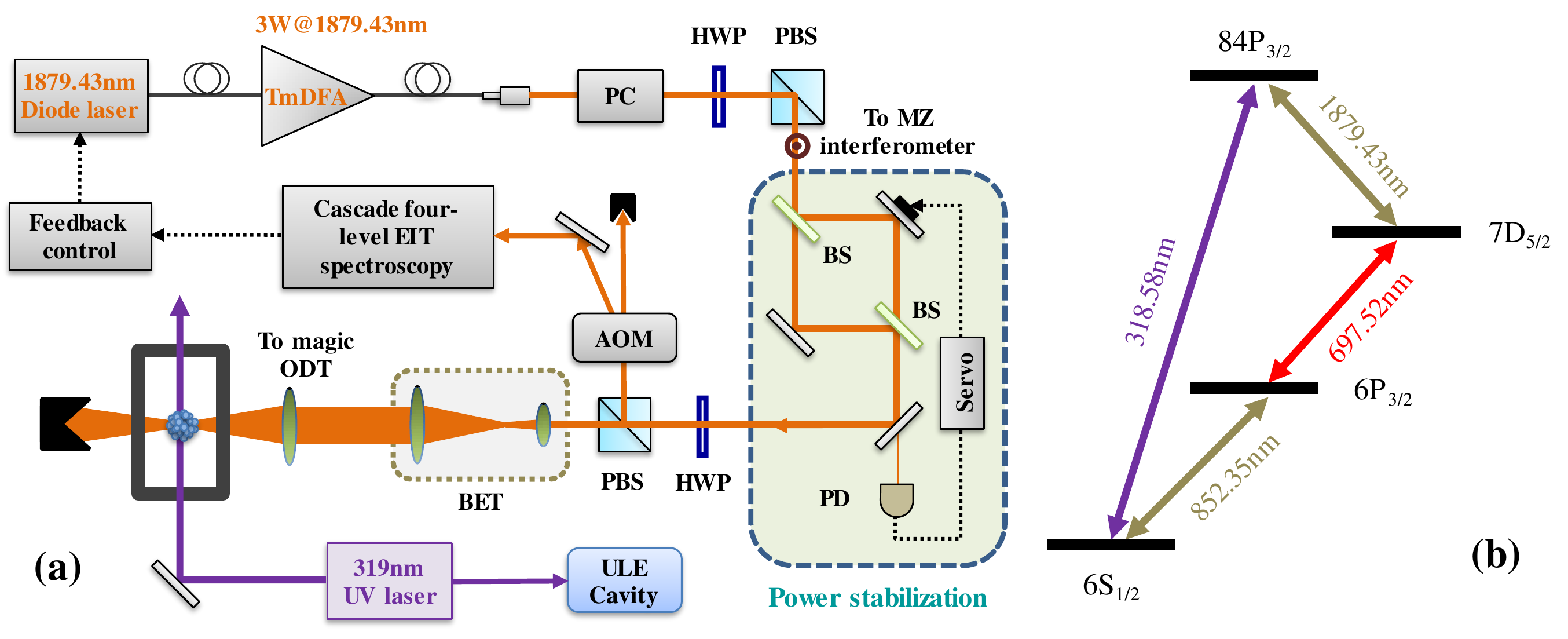}
} \vspace{-0.1in} %\setlength{\columnwidth}{3.2in}
%\centerline{
%}
\caption{(a) Schematic diagram of the 1879.43-nm MOPA laser system. TmDFA, thulium-doped fiber amplifier; PC, polarization controller; HWP, half-wave plate; BS, beam splitter; PBS, polarization beam splitter cube; PD, photodiode detector; AOM, acoustic-optical modulator; BET, beam expanding telescope. (b) The relevant energy levels for single-photon and cascade three-photon transitions.}
\label{Fig 5}
\vspace{0.0in}
\end{figure}

Since the magic detuning of the ODT laser relative to $84P_{3/2}(\vert m_j\vert = 1/2) \leftrightarrow 7D_{5/2}(\vert m_j\vert = 1/2)$ auxiliary transition is about +182 MHz, the frequency stability and continuous tunability of 1879.43-nm laser beam are important for implementation of the ODT. After Mach-Zehnder interferometer, the 1879.43-nm beam is split into two parts by a half-wave plate (HWP) and a polarization beam splitter (PBS) cube. The main portion of the 1879.43-nm beam is used to form the magic ODT. A small fraction is offset-locked to the $84P_{3/2}(\vert m_j\vert = 1/2) \leftrightarrow 7D_{5/2}(\vert m_j\vert = 1/2)$ auxiliary transition by the three-photon (852.35 nm + 697.52 nm + 1879.43 nm) ladder-type four-level electromagnetically induced transparency spectrum (as shown in Fig. 5(b)), which is similar to Ref. \cite{carr2012} and an acoustic-optical modulator (AOM). Therefore, the frequency detuning of the 1879.43-nm beam relative to the auxiliary transition can be controlled by driving frequency of AOM.

To confine cold Cs atoms in the 1879.43-nm magic ODT with a trap potential depth of $\sim$ 1 mK, if the focused beam waist's diameter is $\sim$ 20 $\mu$m, the required optical power will be $\sim$ 1.4 W. In addition, we also estimate the typical size of Rydberg atom, which is proportional to ${(n-{\delta_{n,l,j}})}^2 a_0$. The size of Cs $84P_{3/2}$ Rydberg atom is $\sim$ 1 $\mu$m, which is much smaller than the focused beam waist's diameter of the magic ODT.

After trapping $6S_{1/2}$ ground-state and $84P_{3/2}$ Rydberg-state Cs cold atoms, we will be able to perform Rydberg-dressing experiment to manipulate the strong long-range interaction between atoms by adjusting the fraction of Rydberg state in the coherent superposition state, which is prepared by controlling the intensity and detuning of 319-nm UV laser. The Rydberg-dressing ground state has the long lifetime of ground state, strong controllable long-range interaction between Rydberg atoms, and large electric polarizability, which are promising for applications in quantum simulation and metrology. Recent experiment has demonstrated Rydberg-dressing enhanced Ramsey interferometer and electrometer with a 1560-nm ODT by off-resonantly coupling $4S_{1/2}$ ground state to $39P_{3/2}$ Rydberg state of $^{39}$K atoms by using a 286-nm UV laser \cite{arias2019}. However, the coherence between the two states is still limited by atomic thermal diffusion. The magic ODT we are carrying out will be helpful to solve this issue.

\section{Conclusion}

In conclusion, we propose a magic ODT, which can solve the issue that conventional ODT normally can not confine the highly-excited Rydberg atoms, by adjusting frequency detuning between the ODT laser and the auxiliary transition connected the desired Rydberg state. The magic ODT can confine both ground-state and Rydberg-state Cs cold atoms, which experience the same potential well. It will greatly improve the repetition rate of experimental timing sequence, and avoid the atomic decoherence caused by thermal diffusion. The magic ODT is very promising for implementing quantum information and quantum metrology experiment.

We have calculated the dynamic polarizabilities of Cs $6S_{1/2}$ ground state and $nS_{1/2}$ and $nP_{3/2}$ Rydberg states by using a sum-over-states method. Then we determine the corresponding magic detuning at which ground state and Rydberg state have identical polarizabilities. Furthermore, we analyze the dissipative mechanisms which affect the coherence between ground state and Rydberg state, and estimate the trapping lifetime of Rydberg atom confined in the magic ODT. For our proposed Cs magic ODT, the detuning of ODT laser ranges from several hundred MHz to several GHz, which will result in a bit higher photon scattering rate. However, the lifetime of Rydberg atom in the magic ODT is still close to its radiative lifetime. It is very important to preserve the coherence between ground state and Rydberg state. Moreover, compared with the magic detuning of $\sim$50 MHz for Rb Rydberg transition in Ref. \cite{li2013}, the magic detuning of Cs atoms we calculated is much larger, so that the photon scattering rate is lower. In addition, when the frequency fluctuation of the ODT laser is about several MHz, the fluctuation of trap potential depth is quite small, which is easy to implement by using of a fiber amplifier boosted a master laser and an ultra-stable ULE optical cavity served as frequency reference. Therefore, the experimental scheme is feasible to perform the higher-fidelity entanglement and quantum gate operations.

\section*{Acknowledgements}

This work is financially supported by the National Key R\&D Program of China (2017YFA0304502), the National Natural Science Foundation of China (11774210, 61875111, 11974226, 61905133, and 61475091), and the Shanxi Provincial 1331 Project for Key Subjects Construction.

\end{document}